\documentclass[pra,twocolumn,preprintnumbers,floatfix]{revtex4}
\usepackage{graphicx}
\usepackage{dcolumn}
\usepackage{bm}
\usepackage{latexsym,epsfig}
\usepackage{graphicx}
\usepackage{verbatim}
\usepackage{comment}
\usepackage{amsmath}
\usepackage{amssymb}
\usepackage{stmaryrd}
\usepackage{color}
\usepackage{epstopdf}
\usepackage{grffile}
\DeclareGraphicsExtensions{.eps}

\newcommand{\beq}{\begin{equation}}
\newcommand{\eeq}{\end{equation}}
\newcommand{\bea}{\begin{eqnarray}}
\newcommand{\eea}{\end{eqnarray}}
\newcommand{\ben}{\begin{eqnarray*}}
\newcommand{\een}{\end{eqnarray*}}
\newcommand{\bfig}{\begin{figure}}
\newcommand{\efig}{\end{figure}}

\begin{document}
\title{Three-body constrained bosons in double-well optical lattice}
\author{Suman Mondal$^1$, Sebastian Greschner$^2$ and Tapan Mishra$^{1}$}
\affiliation{$^1$Department of Physics, Indian Institute of Technology, Guwahati-781039, India}
\affiliation{$^2$ Department of Quantum Matter Physics, University of Geneva, 1211 Geneva, Switzerland}

\date{\today}

\begin{abstract}
We analyse the ground-state properties of three-body constrained bosons in a one dimensional optical lattice 
with staggered hoppings analogous to the double well optical lattice. By considering attractive and 
repulsive on-site interactions between the bosons, 
we obtain the phase diagram which exhibits various quantum phases. Due to the double-well geometry and three-body constraint 
several gapped phases such as the Mott insulators and dimer/bond-order phases emerge at commensurate 
densities in the repulsive interaction regime. Attractive interaction leads to the pair formation which leads to the 
pair bond order phase at unit filling which resembles the valence-bond 
solid phase of composite bosonic pairs. At incommensurate densities we see the signatures of the gapless pair superfluid phase. 
\end{abstract}





\maketitle
\section{Introduction}
Ultracold atoms in optical lattices with tunable interactions and lattice parameters have opened up a wide area of research in recent 
years. The significant progress both in theoretical and experimental fronts have uncovered a wealth of new physics which was impossible to 
achieve in the conventional solid state systems. The path breaking observation of the superfluid~(SF) to Mott insulator~(MI) 
transition~\cite{bloch} following its theoretical prediction~\cite{jaksch}
in a system of ultracold bosons in optical lattice have paved the path to simulate complex quantum many-body physics~\cite{lewenstein}. 
The exquisite control over the interactions and lattice geometries are the key to achieve such interesting physics. 
A new frontier of research have evolved with the construction of two color superlattices which is an array of double-well 
potentials~\cite{danshita6,danshita7,danshita8,danshita9,danshita1,danshita2,Silva2016,bloch}. 
Recent experiments on systems of ultracold atoms in these double wells have led 
to various interesting phenomena in condensed matter physics and also in atom interferometry~\cite{Sabby2007} and quantum information~\cite{Marco2007}. 
Particularly in one dimension these double-well lattices exhibits interesting 
properties due to the staggered or dimerized hopping amplitudes. 
The presence of this staggered hopping resembles these system to the inversion symmetric 
Su-Schrieffer-Heeger(SSH) model for fermions~\cite{ssh} which possess interesting topological features characterized
by the Zak phase~\cite{Ryu2002,Zak1989,Delplace2011,Lang2012}. The model has been generalized to interacting 
fermions~\cite{Manmana2012,Yoshida2014} and bosons~\cite{Grusdt2013,Silva2016} and explored in recent experiments~\cite{Atala2013,Schweizer2016,Lohse2016, Takahashi2016pumping,Schweizer2016,Browaeyes2018ssh}.

\begin{figure}[!b]
\begin{center}
\includegraphics[width=0.8\columnwidth]{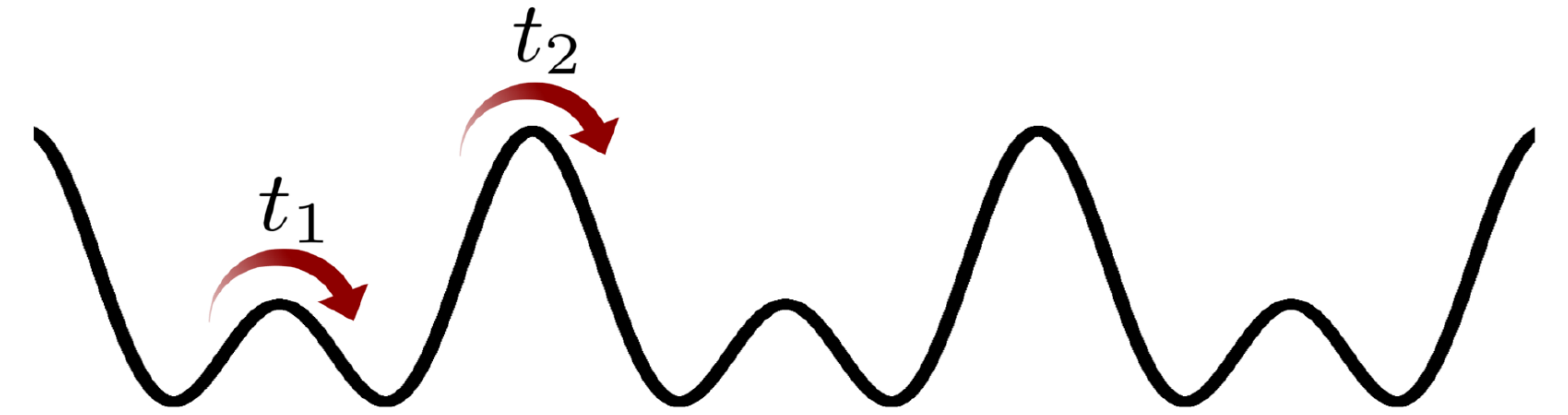}
 \end{center}
\caption{One dimensional double well optical superlattice with staggered hoppings $t_1 > t_2$ .}
\label{fig:lattice}
\end{figure}

On the other hand the experimental observation of local higher order interactions in optical lattices 
have opened up a new direction to simulate quantum phase transitions in the 
presence of multi-body interactions~\cite{wills}. 
Several interesting 
ideas have been proposed to engineer and tune such interactions in optical lattices~\cite{petrov1,petrov2,tiesinga,daley,sansone}.
One of such example is the possibility to create a situation where 
the three-body interaction can become extremely large. Under this circumstance the bosons experience 
three-body hardcore constraint~\cite{baranovprl}, which prohibits 
more than two atoms to occupy a single lattice site. This condition facilitates to explore the physics of attractive Bose-Hubbard model which 
otherwise leads to collapse of atoms onto a single site. Many novel scenarios have been investigated recently by considering three-body 
constrained bosons in optical lattices~\cite{baranovprl,wessel,zhang,sowinski1,singh1,singh2,sebastian1,Hincapie2016,Silva2014,Silva2012}. The 
physics which is manifested by the two body interaction along with the large three body repulsion is one of the simplest 
problem to understand although it has interesting physical implications. In such a scenario the system exhibits
the usual SF-MI phase transition for repulsive 
on-site interaction and for attractive interactions there exists a superfluid to pair superfluid~(PSF) phase transition~\cite{wessel}. 

In this paper we consider a system of three-body constrained bosons in a double well 
optical superlattice in one dimension which resembles the SSH type model as shown in Fig.~\ref{fig:lattice}. 
The presence of double well potentials creates a 
situation with staggered hopping amplitudes. This superlattice geometry can be created by superimposing 
two lattices with one lattice having double the period of the other.



The physics of ultracold bosons in double well optical lattice can be explained by a modified Bose-Hubbard model 
with staggered hopping amplitudes(bosonic SSH model) which is given as; 
\begin{align}
\mathcal{H} &= - t_1 \sum_{i~\in~ odd} (a_{i}^{\dagger}a_{i+1}^{\phantom \dagger} + \text{H.c.})\nonumber\\
&- t_2\sum_{i~\in~ even}(a_{i}^{\dagger}a_{i+1}^{\phantom \dagger}+\text{H.c.}) 
 + \frac{U}{2}\sum_{i}  n_i(n_i-1)
\label{eq:ham}
\end{align}
where $a_i^{\dagger}$ and $a_i^{\phantom \dagger}$ are the creation and annihilation operators
for bosons at site $i$ and $n_i=a_i^{\dagger}a_i^{\phantom \dagger}$ is the number operator 
at site $i$. $t_1$ and $t_2$ are the tunneling rates from odd and even sites respectively (compare Fig.~\ref{fig:lattice}). 
The onsite contact interactions are characterized by the term $U$. The bosons in the lattice enjoy three-body 
constraint i.e. $(a_i^{\dagger})^3=0$. 

At half filling the single particle spectrum of Model(\ref{eq:ham}) exibits gap for any imbalance in hopping between 
the unit cells $t_1\neq t_2$~\cite{ssh}. Hence, the the ground state is a dimer phase or bond order~(BO) phase for 
spin polarized  fermions or bosons with very large on-site interactions~(hardcore bosons). The presence of three-body constraint may lead to 
interesting phenomena in such dimerized lattice which will be the topic of this paper. In particular we study the interplay between pairing of particles 
and dimerization which gives rise to the emergence of a pair-bond-ordered phases with a sharp crossover to the bond-ordered phase.
We assume $t_1=1$ (unless stated otherwise) which makes other physical quantities dimensionless.  
Ground state properties of this system is analysed using the density matrix renormalization group~(DMRG) method. We consider system 
sizes up to $160$ sites and retaining up to $800$ density matrix eigenstates. 

The rest of the paper is arranged as follows. In Sec.~II we discuss the limiting cases focusing on 
hardcore bosons in the optical lattice made up of series of double wells and the physics in the limit of isolated double wells. 
In Sec.~III, 
we present the general phase diagram 
with different values of twobody onsite interactions. In Section~IV we give a brief summary of the work. 


\section{Limiting cases}

We begin our discussion with a short summary of the properties of two analytically solvable limits of model~(\ref{eq:ham}). These 
analysis will help in understanding the physics of the system discussed in the paper.

\subsection{$U=\infty$ limit}

\begin{figure}[!t]
  \centering
  \includegraphics[width=\linewidth]{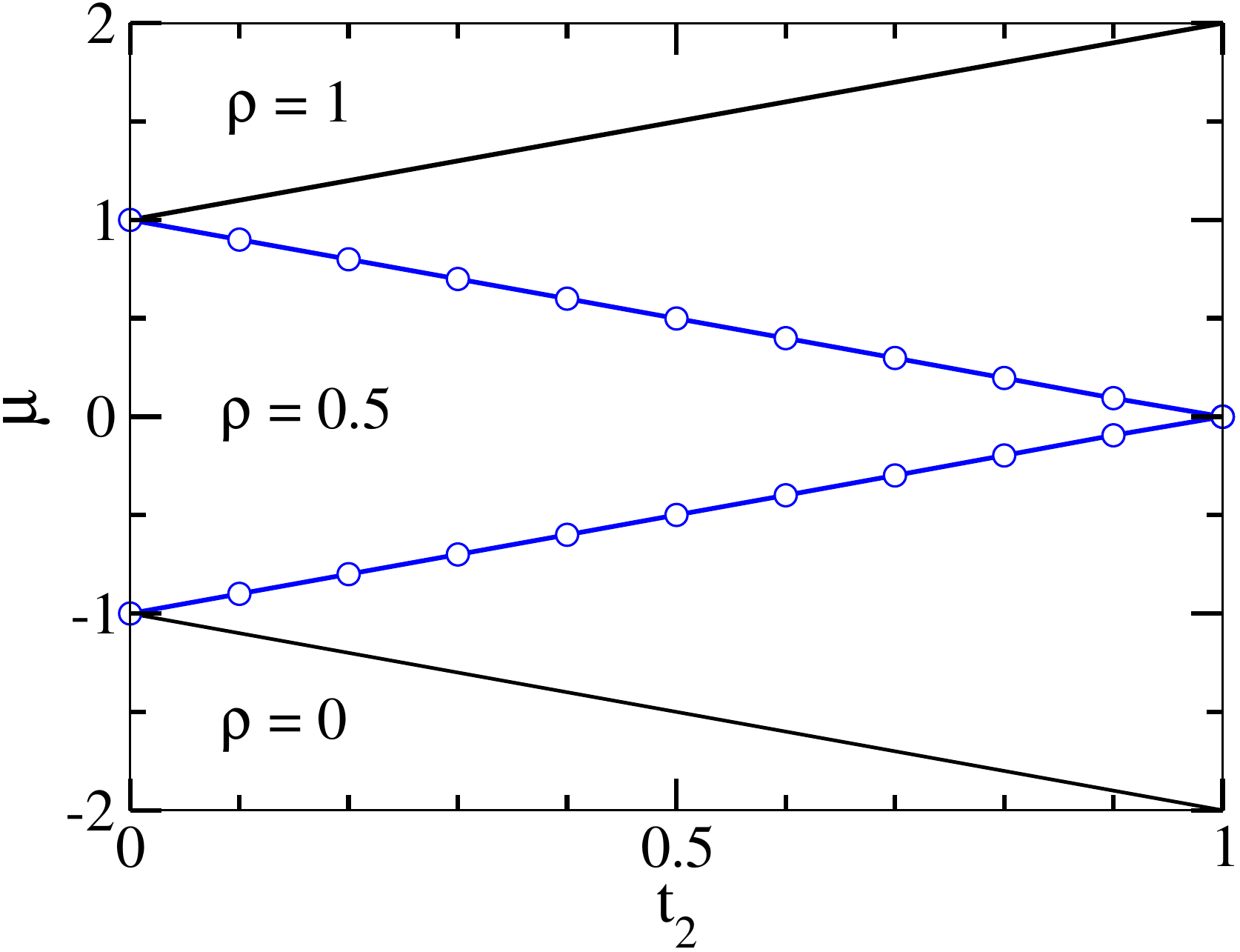}
    \caption{(Color online)Phase diagram of hardcore bosons in the $\mu$ - $t_2$ plane with $t_1=1.0$.
The points correspond to extrapolated DMRG data with $L=300$, which accurately lie on top of the analytical curves (solid lines).
    The region included by the blue circles is the gapped bond-ordered~(BO) 
    phase at half filling. The gap opens up for any finite dimerization $t_2/t_1$}. The black curves represent the empty and full states.
    \label{fig:hcpd}
\end{figure}

In the limit of large interactions $U\rightarrow \infty$, the bosons are hardcore in nature and in this limit 
the Model~(\ref{eq:ham}), after a Jordan-Wigner transformation to free fermions, can be considered as the topological 
SSH model as mentioned before. Due to the staggered hopping amplitudes, the SSH model at half filling dimerizes naturally due to the Peierls instability 
and one gets dimerized phase of 
bosons. In this phase a single boson lives in one of the unit cell composed of two 
lattice sites in the double-well with larger 
hopping strength. Doping away from half filling breaks this dimer ordering and a critical SF phase appears. 
This gapped phase is called the BO phase. 
Note, that this BO phase, stabilized due a spin-Peierls like mechanism~\cite{GiamarchiBook}, does not 
exhibit spontaneous symmetry breaking and the BO-order is induced due to the explicitly broken translational symmetry of the model. 
In the literature this phase is, hence, also called fractional Mott-insulator phase or similar phase~\cite{Grusdt2013}.
The phase diagram of such system is shown in Fig.~\ref{fig:hcpd} as a function of $t_2$ and the filling. 
The gapped phases are characterized by the finite single particle gaps which is defined as 
\begin{equation}
 E_G=\mu^+ - \mu^-,
 \label{eq:gap}
\end{equation}
where $\mu^+$ and $\mu^-$ are the chemical potentials. As it can be seen from the 
phase diagram, any finite hopping imbalance leads to the gapped phase which is a bond-ordered~(BO) phase and this phenomenon 
is also evident from the single particle spectrum.  

\subsection{Isolated Double Wells}
\label{sec:doublewell}
\begin{figure}[!t]
  \centering
  \includegraphics[width=\linewidth]{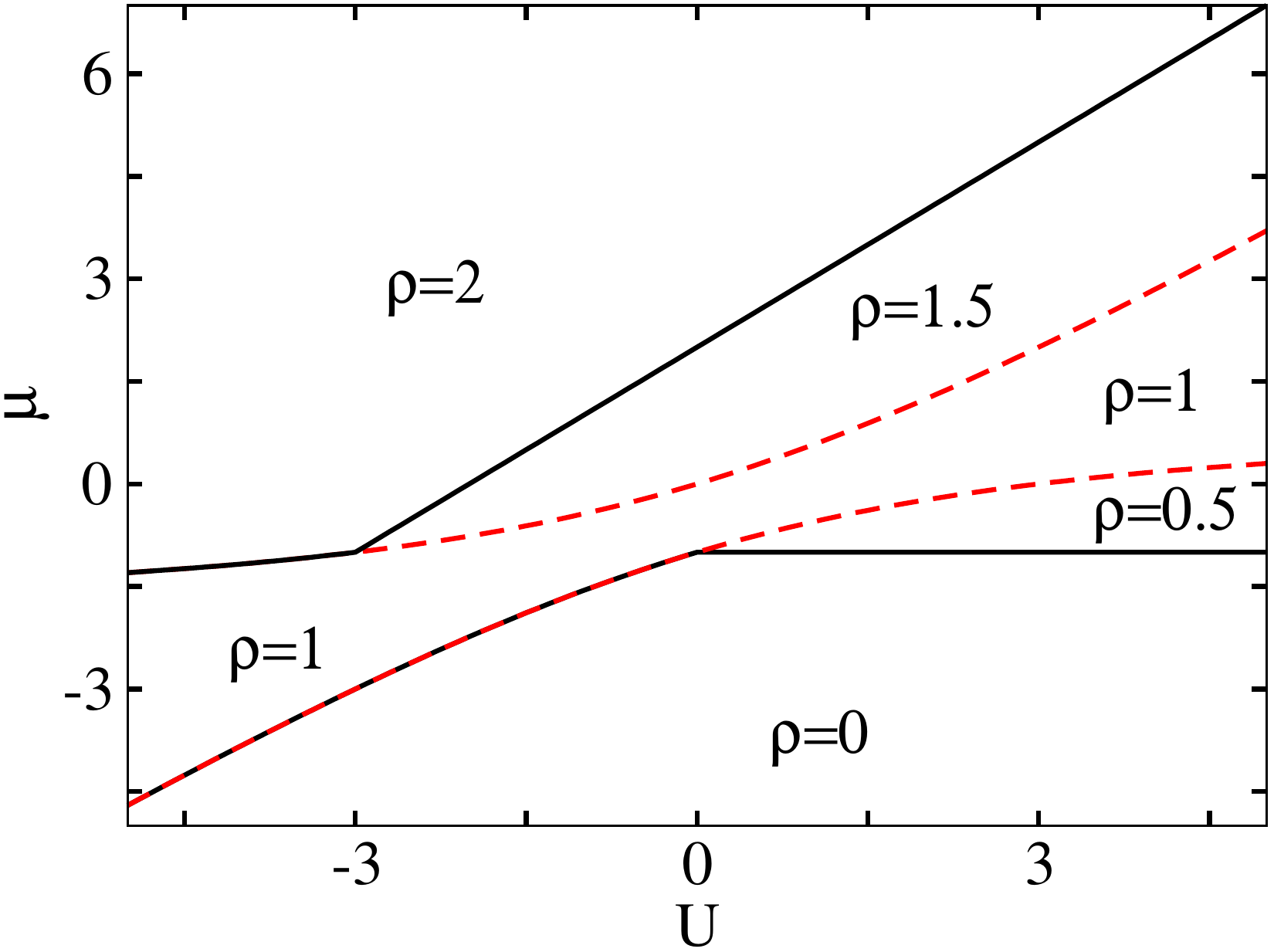}
    \caption{(Color online)Phase diagram of isolated doublewells showing three gapped phases at $\rho= 0.5$, $1$ and $1.5$. 
    }
    \label{fig:pddoublewell}
\end{figure}

It is also instructive to discuss the model for the case of trivially
disconnected double-wells, corresponding to the case $t_2=0$. From this limit one can conveniently explain the finite hopping
case which is the topic of interest of the paper. In this limit 
the model Hamiltonian can be readily diagonalized for a fixed particle number:
In the $n=1$ ($n=3$) sector two eigen-energies $\pm t_1$ ($U \pm 2 t_1$)
are found. For $n=2$, the eigenvalues are given by $U$ and
$\frac{1}{2} \left(U \pm \sqrt{16 t_1^2+U^2}\right)$.
With these eigenvalues in a grand-canonical ensemble three gapped
phases at fillings $\rho=0.5$, $1$ and $1.5$ can be observed as shown in Fig.~\ref{fig:pddoublewell} in the strong dimerization limit. 
In this limit the gap at unit filling is given by $E_G = -3 t_1
+ \sqrt{16 t_1^2 + U^2}$.

The ground state in the $n=1$ sector is given by $|\psi_1\rangle = \psi_{20}
|2,0\rangle +  \psi_{11} |1,1\rangle + \psi_{02} |0,2\rangle$. Here
$|n_1, n_2\rangle$ denotes a Fock-state basis of the isolated double
well and $\psi_{02}=\psi_{20}=2/\sqrt{16t_1^2+2U\varepsilon}$,
$\psi_{11}=2\varepsilon/\sqrt{16t_1^2+2U\varepsilon}$ where
$\varepsilon=U/2 + \sqrt{4 t_1^2 + U^2/4}$.
For $U\to \infty$ this results in a MI like state $\sim
|11\rangle$ and a dimer of pairs $|20\rangle + |02\rangle$ (pair-bond-ordered or PBO phase) for strong
attractive interactions with a smooth crossover between both regimes.
Indeed, the decoupled double well ground state for $U=0$ resembles a superposition of MI and PBO states. 
The features of this interesting many-body state for finite hopping $t_1,t_2\neq 0$ will be studied in Sections~(III).

A finite small hopping $0<t_2\ll 1$ will couple the double wells and
allow for a melting of the gapped phases due to the energy gain by
delocalization of excitations and stabilize superfluids separating the
gapped phases. This process may be understood as well from a effective
model of coupled dimer-states such as recently discussed in
Refs.~\cite{DiLiberto2016, DiLiberto2017}.
\begin{figure}[!b]
  \centering
  \includegraphics[width=\linewidth]{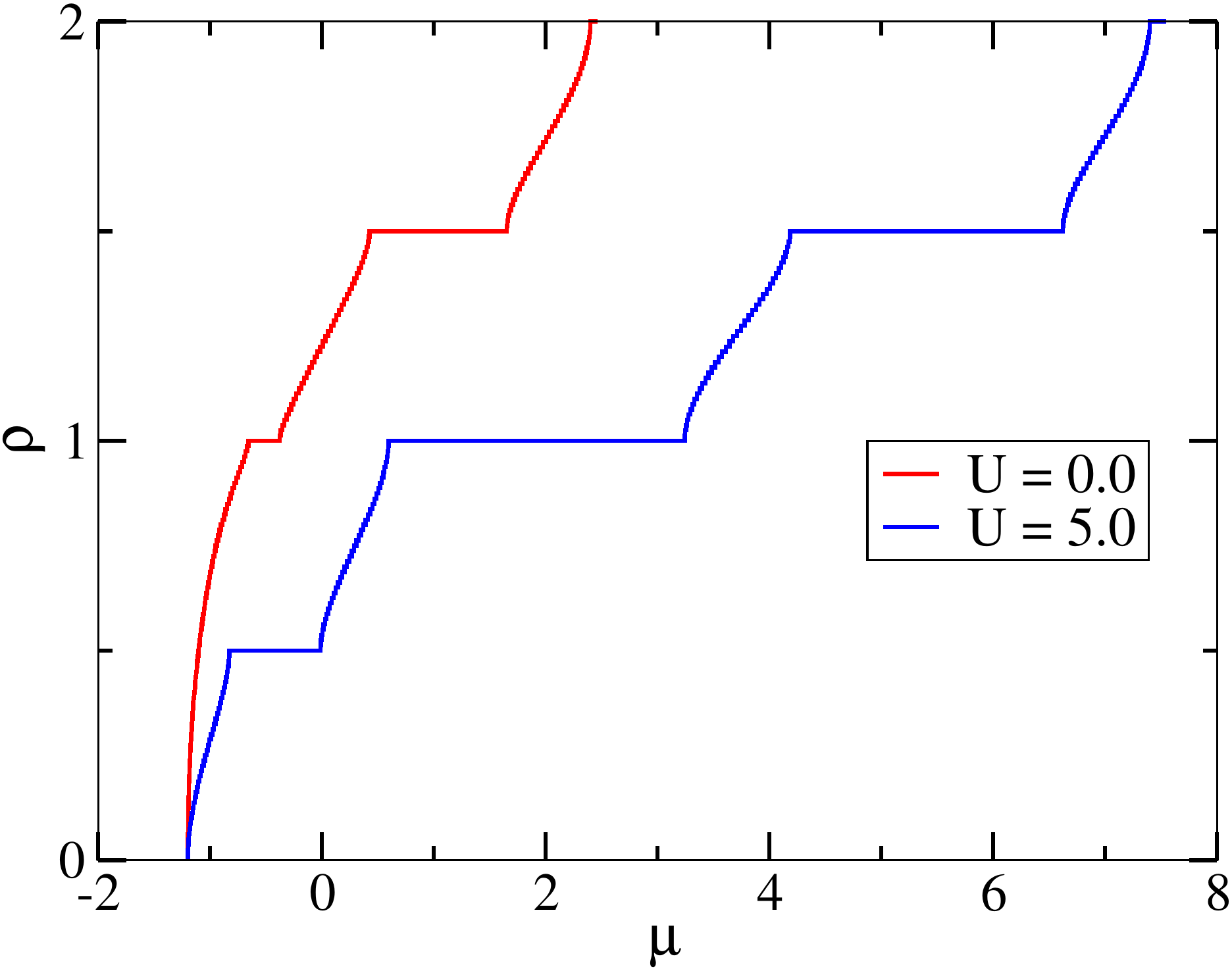}
    \caption{(Color online)Figure shows the behaviour of $\rho$ with respect to $\mu$ for $U=0$ (left red curve) and $U=5$ (right blue curve) for $t_2=0.2$. Plateaus indicate the gapped phases. 
    }
    \label{fig:rhomu}
\end{figure}


\section{General Phase Diagram}
In this section we discuss the most general model with finite onsite interaction and finite hoppings. To start with we consider the case of vanishing interaction and 
then we switch on interaction to see the effect of dimerised hopping on the physics of the system. As our main finding we discuss the emergence and crossover between the various gapped 
phases such as bond-order, pair-bond-order and the Mott-insulator phases. We show how these phases may be characterized by their different measurables such as the dimerization 
and their characteristic parity order.

\subsection{Vanishing two-body interaction($U=0$)}
In the limit of vanishing interactions for a softcore boson 
without the three-body hardcore constraint one expects an SF phase even for very strong hopping imbalance. 
In the presence of interaction 
the physics of the system is governed by 
the competition between the hopping amplitudes and the onsite interactions which leads to the non-trivial gapped phases at 
intermediate half integer filling apart from the gapped MI phases~\cite{Grusdt2013} as a function of interaction $U$. 
A similar feature is also present in the case of usual two color superlattice potential where the SF phase becomes gapped 
MI phases at half integer and integer fillings for strong interactions~\cite{manpreetsuplat}. 
\begin{figure}[!t]
  \centering
  \includegraphics[width=\linewidth]{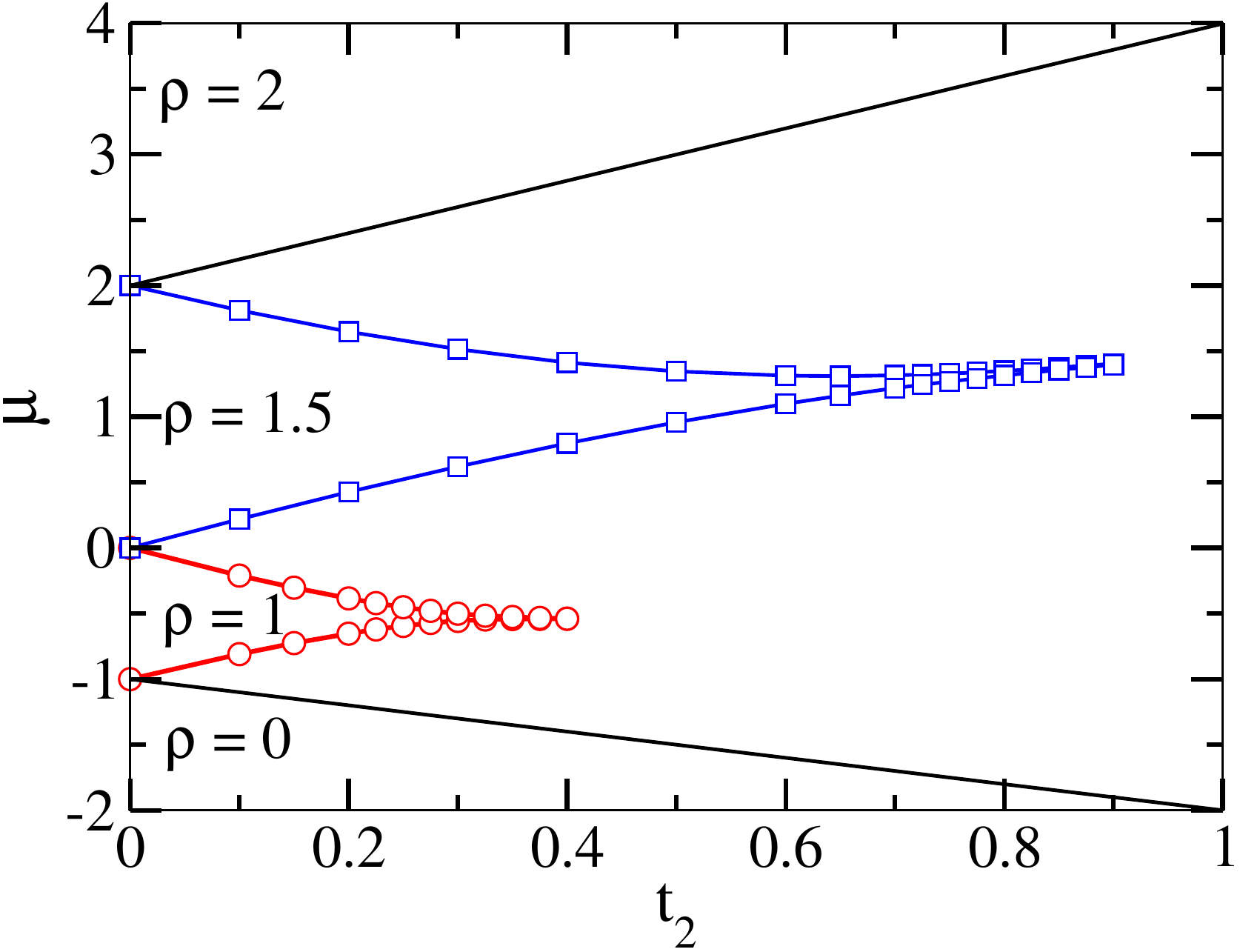}
    \caption{(Color online)The phase diagram for three-body constrained bosons as a function of $t_2$ for $U=0$ obtained from polynomial extrapolation of gap. 
    The BO phase at $\rho=1$ and $\rho=1.5$ appear at $t_2 \sim 0.4$ and $t_2 \sim 0.9$ respectively.
     }
    \label{fig:nonint}
\end{figure}
The situation however, is different in the case of three-body constrained 
bosons where a maximum of two bosons can occupy a single lattice site. 
Due to the effect of double well superlattice the motion of particles is restricted to one unit cell. 
Interestingly, in such a scenario two different gapped phases arise at $\rho=1$ and $\rho=1.5$ after some 
critical values of $t_2$. The gap in the system can be seen as the finite plateaus in the $\rho$ vs $\mu$ diagram as shown in Fig.~\ref{fig:rhomu}.
The phase diagram corresponding to this scenario is depicted in Fig.~\ref{fig:nonint} where the gapped phases at $\rho=1$ 
and $\rho=1.5$ appear at $t_2 \sim 0.4$ and $t_2\sim 0.9$ respectively.
 The gapped BO regions for $\rho=1$ and $\rho=1.5$ are bounded by red circles and blue squares respectively. 
 The black lines correspond to the empty and full states. 
The remaining part of the phase diagram is the SF phase. 



\subsection{Finite U and $t_2=0.2$ case}
\begin{figure}[!t]
  \centering
  \includegraphics[width=\linewidth]{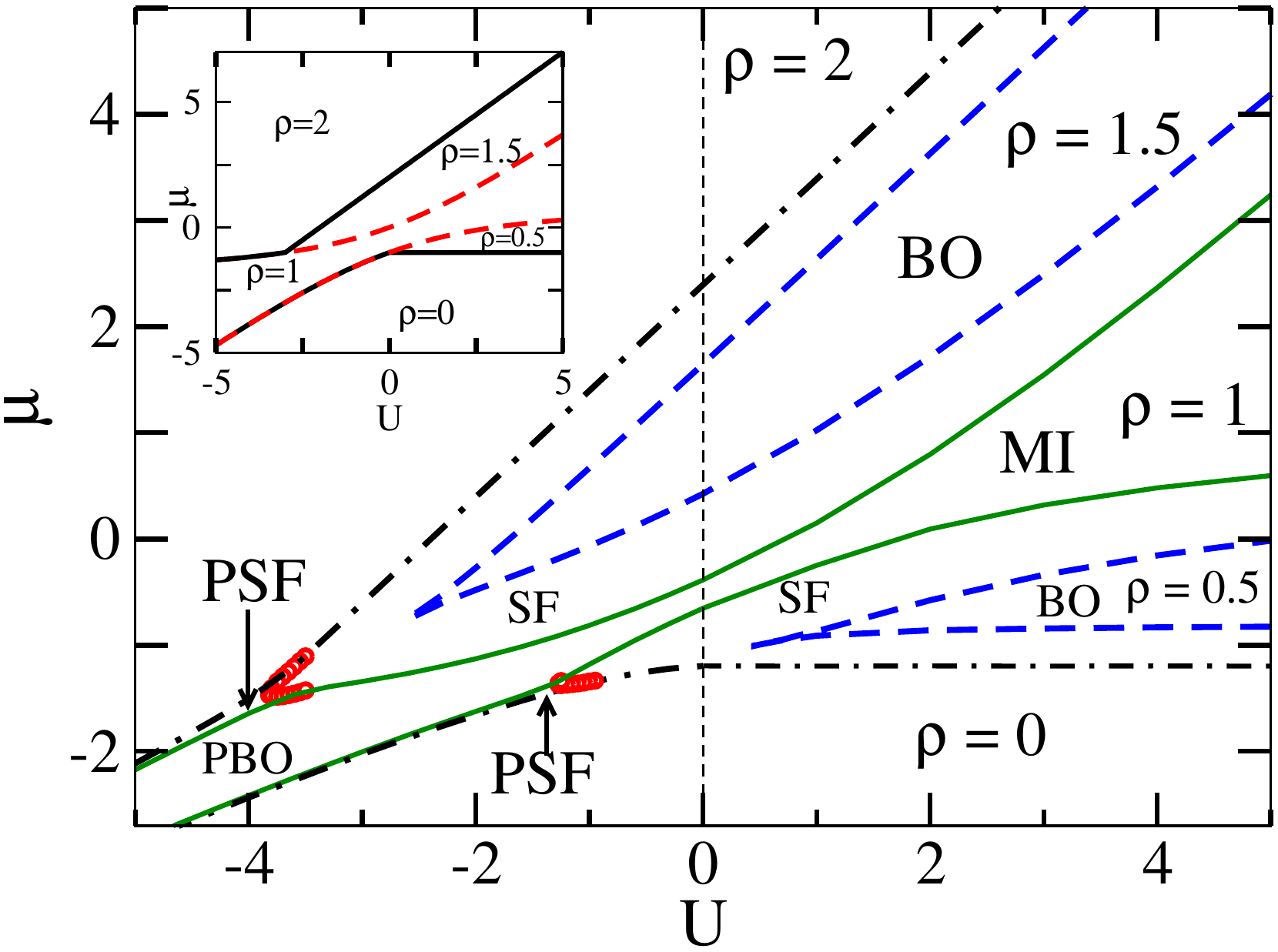}
    \caption{(Color online)Phase diagrams for $t_2=0.2$. The region bounded by the green curves are the gapped phases at $\rho=1$(middle) which consists of the MI(PBO) phases in the strong 
    repulsive(attractive) regimes. The regions bounded by the blue dashed curves are the gapped BO phases at $\rho=0.5$(upper) and $\rho=1.5$(lower). 
    On the attractive side the PSF is separated from the SF phase by the red circles. (Inset) Shows the phase diagram in the limit of isolated 
    double-wells for comparison.}
    \label{fig:fullpd1}
\end{figure}

As the system is already in the gapped BO phase for $U=0$ at commensurate densities except at $\rho=0.5$, it is interesting to see the 
effect of interactions on the ground state of the system. The phase diagram for this case is shown in Fig.~\ref{fig:fullpd1}. As we move away from the $U=0$ limit along the positive $U$ axis  
the gapped phases grow as can be seen from the enlargement of the plateaus in the $\rho$ vs $\mu$ plot for $U=5$ shown in Fig.~\ref{fig:rhomu}. 
The gapped phases at $\rho=0.5$ and $\rho=1.5$ are depicted
 by the region bounded by the blue dashed curves and the one at $\rho=1$ is bounded by the green solid 
 curve in the phase diagram of Fig.~\ref{fig:fullpd1}.
At $\rho=0.5$, the gap appears after a critical point  $U\gtrsim 0.4$ leading to the BO phase. 
As anticipated in the discussion of the decoupled double-well case, the excitation gap at  $\rho=1$ remains finite for all $U$ even for a 
small $t_2>0$ leading to a smooth cross-over from the pair-bond-ordered~(PBO) phase to the MI phase through the BO phase 
where every site is occupied by one atom due to large onsite repulsion. 
The strong onsite repulsion disfavours the dimerization and prohibits two particles occupying a single site. 
For $\rho=1.5$, the system remains in the BO phase which becomes wider as a function of $U$. 
This BO phase is similar to the one for the hardcore bosons at $\rho=0.5$ as discussed before.
 The boundaries for the gapped phases are obtained by computing the chemical potentials $\mu^+$ and $\mu^-$ and 
 extrapolating them to thermodynamic limit using system sizes of $L=20,~40 $ and $80$. 

\begin{figure}[!b]
  \centering
  \includegraphics[width=\linewidth]{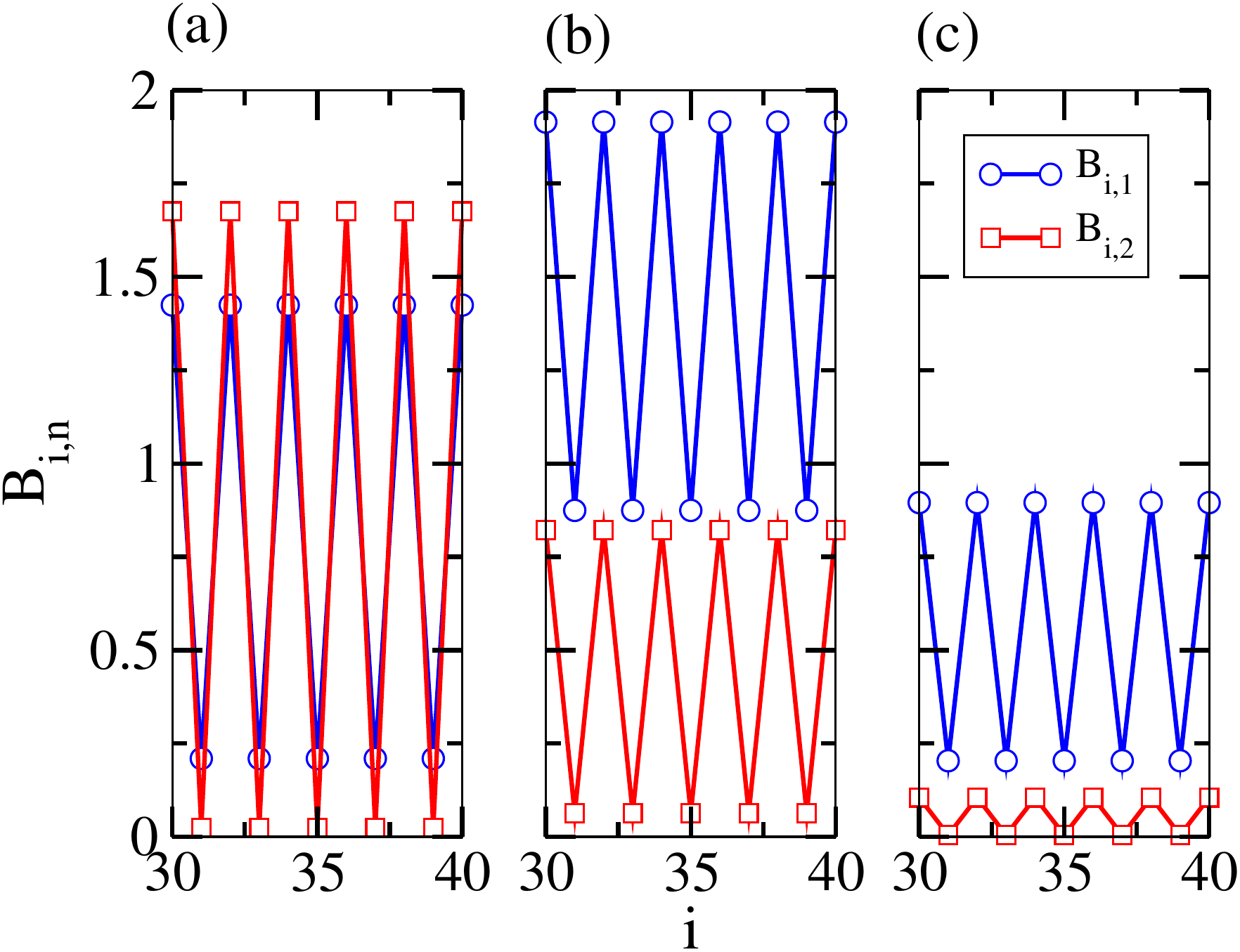}
    \caption{(Color online)Bond operator $B_{i,n}$ are shown for the BO(blue circles) and PBO(red squares) phases for $\rho=1$ and $t_2=0.2$. (a)$B_{i,n}$ for $U=-4$. (b)$B_{i,n}$ for $U=0$. (c) $B_{i,n}$ for $U=8$. }
    \label{fig:bo_one}
\end{figure}

\begin{figure}[!t]
  \centering
  \includegraphics[width=\linewidth]{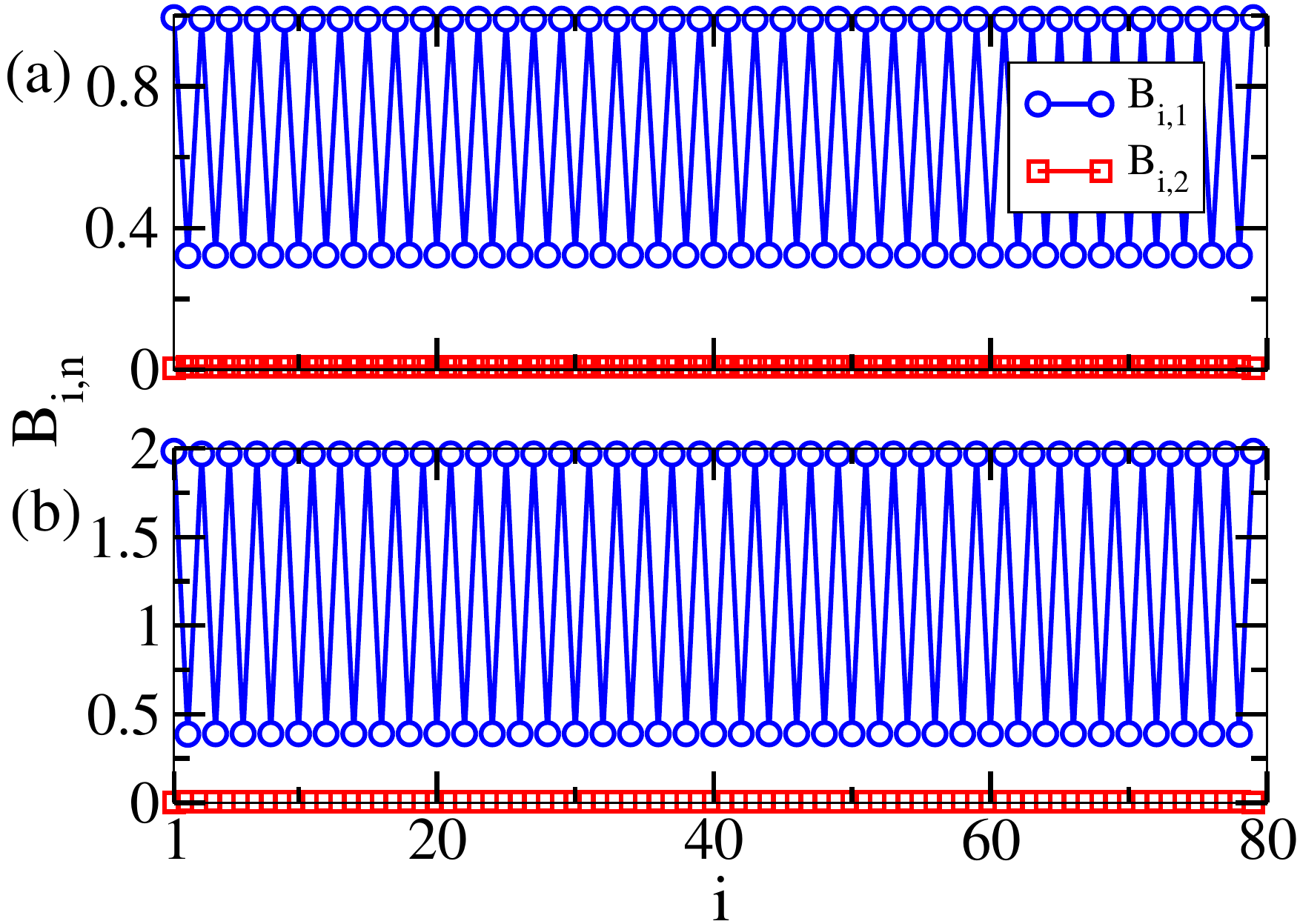}
    \caption{(Color online)Bond operator $B_{i,n}$ are shown for the BO(blue circles) and PBO(red squares) phases for $t_2=0.2$. (a)$B_{i,n}$ for $U=4$ and $\rho=0.5$. (b)$B_{i,n}$ for $U=2$ and $\rho=1.5$.}
    \label{fig:bo_half}
\end{figure}

 The signatures of BO and PBO phases can be seen by plotting the bond operator 
 \begin{equation}
  B_{i,n} = \langle {(a_i^{\dag}})^n (a_{i+1})^n + H.c\rangle
 \end{equation}
 for 
 different bonds. Here the exponent $n = 1$ and $n = 2$  for the BO and the PBO phases respectively. In Fig.\ref{fig:bo_one} we plot $B_{i,n}$ for unit filling which show finite oscillations 
 in the BO and PBO phases. The calculations are done by taking $80$ sites and in the figure we show only the central part of the system. Fig.\ref{fig:bo_one}(a), Fig.\ref{fig:bo_one}(b) and 
 Fig.\ref{fig:bo_one}(c)
 shows the values of $B_{i,n}$  for 
 $U=-4, 0$ and $8$ respectively. The strong oscillation of PBO operator compared to the BO operator for $U = -4$ shows that the system is dominantly in 
 the PBO phase. Also it can be seen that for $U=8$ the oscillation of $B_{i,n}$ has decreased drastically due to the MI phase.
 Similarly, in Fig.\ref{fig:bo_half}(a) and Fig.\ref{fig:bo_half}(b) we plot the value of $B_{i, n}$ for $\rho = 0.5$ and $\rho = 1.5$ respectively for 
 the repulsive values of $U$ where the system is in the BO phase.

Further we obtain the signature of the BO phases by computing the 
bond-bond correlation function and the related structure factor which is given as;
\begin{equation}
 S_{BO}(k)=\frac{1}{L^2}\sum_{i,j}e^{ikr}\langle B_iB_j\rangle,
 \label{eq:bostr}
\end{equation}

\begin{figure}[!b]
  \centering
  \includegraphics[width=\linewidth]{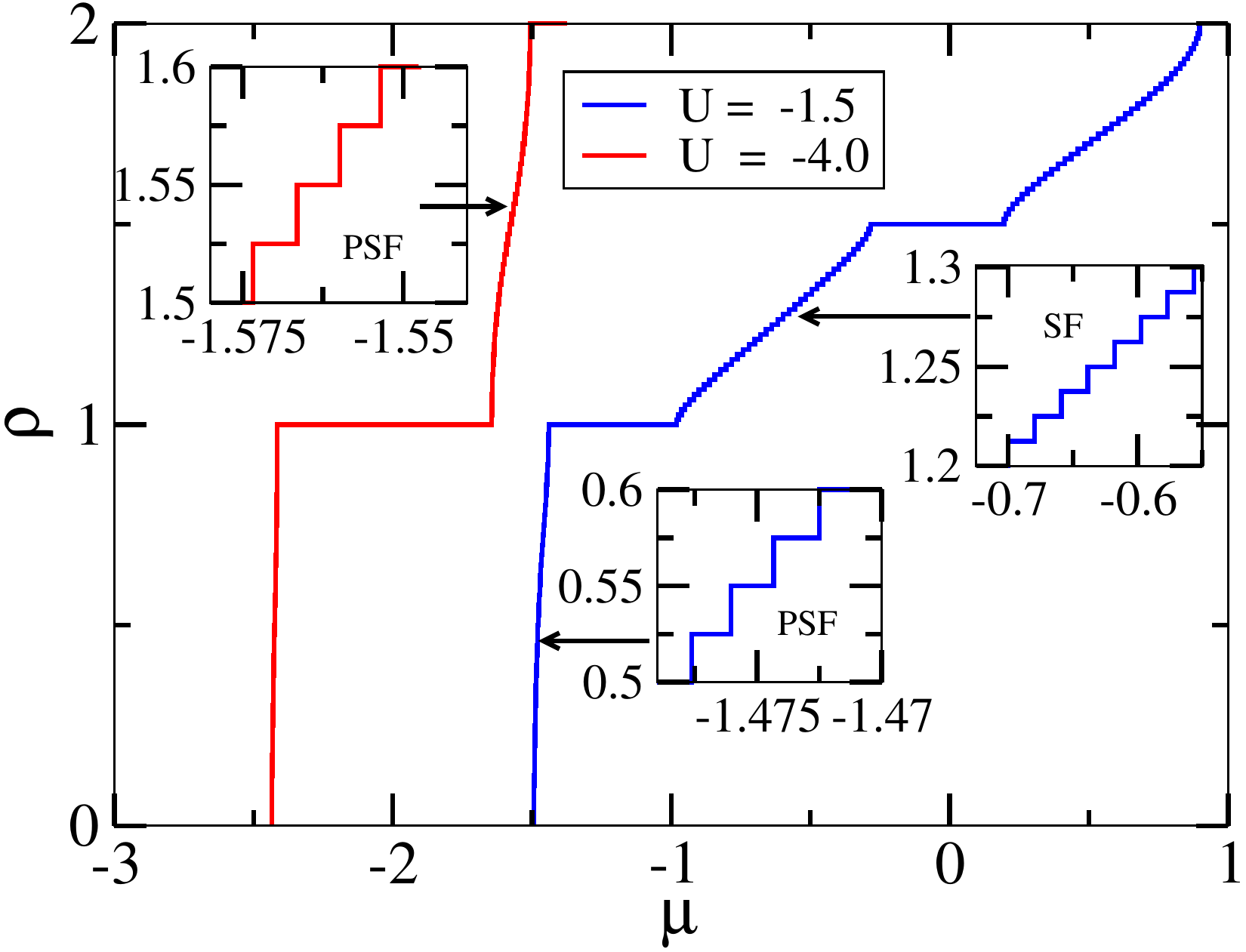}
    \caption{(Color online)Behaviour of $\rho$ with respect to $\mu$ for $U=-4.0$ (left red curve)  and $U=-1.5$ (right blue curve) at $t_2=0.2$. 
    The insets show the enlarged regions of the SF and PSF phases where the density jumps in steps of one and two particles respectively. }
    \label{fig:rhomub}
\end{figure}
where $r=|i-j|$ is the distance. In the BO phase the quantity $B_i$ oscillates in alternate bonds 
and the structure factor exhibits a finite peak at the zone boundaries. 
It is to be noted that the BO phases which appear in the phase diagram are 
 not the true BO phase as lattice translational 
 symmetry is not spontaneously broken. However, the signature is similar to the BO phase whose qualitative feature can be seen from 
the bond order structure factor. 

\begin{figure}[!b]
  \centering
  \includegraphics[width=\linewidth]{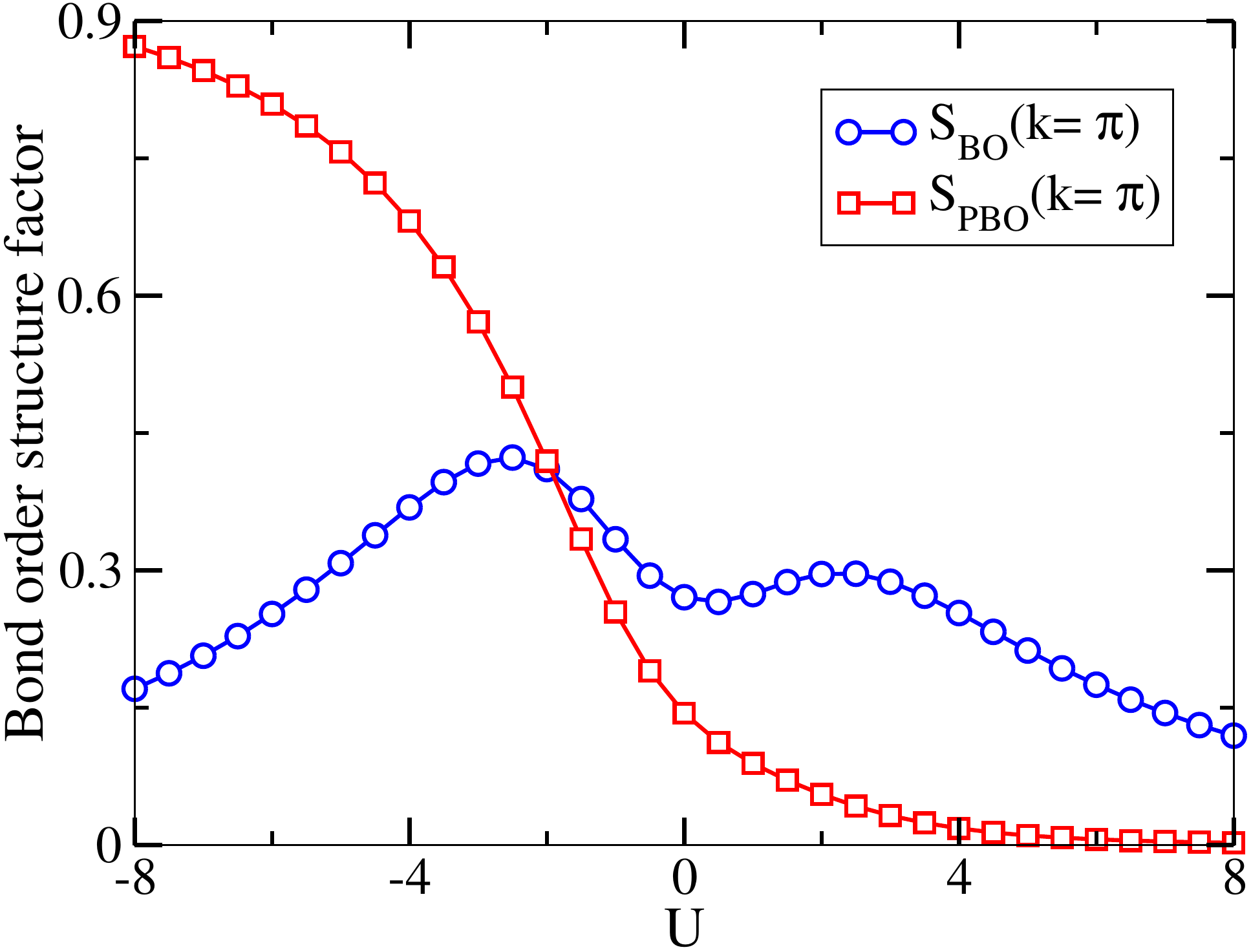}
    \caption{(Color online)The extrapolated values of bond order structure factor for single particle (blue circles) and pairs (red squares) for different values of $U$ at $\rho=1$ (see text for details). Here $t_2=0.2$.}
    \label{fig:bostr}
\end{figure}

 The situation becomes interesting in the attractive regime. Because of the three-body constraint the system is stable against collapse and due to the attractive interaction, the particles start to form pairs. 
 While there is no gapped phase at $\rho=0.5$ in this side of the phase diagram, the gap at $\rho=1.5$ remains finite up to some finite values of $U$ and then closes after a critical point
 of {$U= -2.6$ }. The closing up of the gap 
is due to the competition between the hopping and attractive interaction which tries to break the dimerization and system becomes a 
superfluid as shown in Fig.~\ref{fig:fullpd1}. However, for $\rho=1$, the gap survives for very large values of $U$ extending up to infinity. 
These features can be seen from the plateaus in the $\rho-\mu$ plot as 
 shown in Fig.~\ref{fig:rhomub} for two different 
 values of {$U=-1.5$ and $U=-4.0$}. For $U=-1.5$, the gaps appear at $\rho=1$ as well as at $\rho=1.5$ where as for $U=-4$, 
 only the gap at $\rho=1$ exists. 
 For sufficiently strong attractive interactions, the particles tend to form pairs and at unit filling it becomes a half 
 filled system of bosonic pairs. It is to be noted that these pairs behave like 
 hardcore bosons due to the three-body constraint. In such a scenario the ground state is similar to the dimerization of 
 the hardcore bosons as discussed in the section for $U=\infty$ case. 
 The gapped phase for large attractive $U$ is the BO phase of pairs which can be called as a PBO phase. 
 However, in the weak interaction regime pair formation is not favoured due to the competition between the interaction and kinetic energy. 
 Therefore, the BO phase which appear at $U=0$ survives up to a finite value of attractive $U$ and then there exist a smooth crossover to the PBO phase as the value of $U$ increases. 

 The characteristic feature of the PBO phase is seen from the PBO structure factor $S_{PBO}(k)$ which is similar to the BO structure factor as 
 defined in Eq.~\eqref{eq:bostr} with $a^\dagger(a)$ replaced by ${a^\dagger} ^2(a^2)$ which is shown in Fig.~\ref{fig:bostr}. It can be seen that the value of $S_{PBO}(k=\pi)$ increases smoothly as the 
 value of 
 $U$ becomes more and more attractive. At the same time the BO structure factor $S_{BO}(k=\pi)$ decreases smoothly after 
 increasing up to a particular value of attractive interaction $U\sim-2.5$. 
 Note that the finite 
 value of $S_{PBO}(k=\pi)$ for repulsive $U$ and small attractive $U$ are due to the finite probability of second order 
 hopping processes in the BO phase. We would like to stress that since the BO phases are 
 the manifestation of the double well geometry of the lattice, the BO structure factor remains finite even in the the MI and PBO phase even in the thermodynamic limit. 
It can be seen from Fig.~\ref{fig:bostr} that on the repulsive interaction side the value of $S_{BO}(\pi)$ 
 smoothly decreases after a particular value of $U$. 
 This signals the crossover to the MI phase.  
 The onset of the MI phase can be understood as in the case of large interaction $U$, 
 the system prefers to accommodate one particle in each site which is also true in the homogeneous 
 lattice systems. This can be seen from the decreasing trend of the curve shown in the Fig.~\ref{fig:bostr}. 

\begin{figure}[!b]
  \centering
  \includegraphics[width=\linewidth]{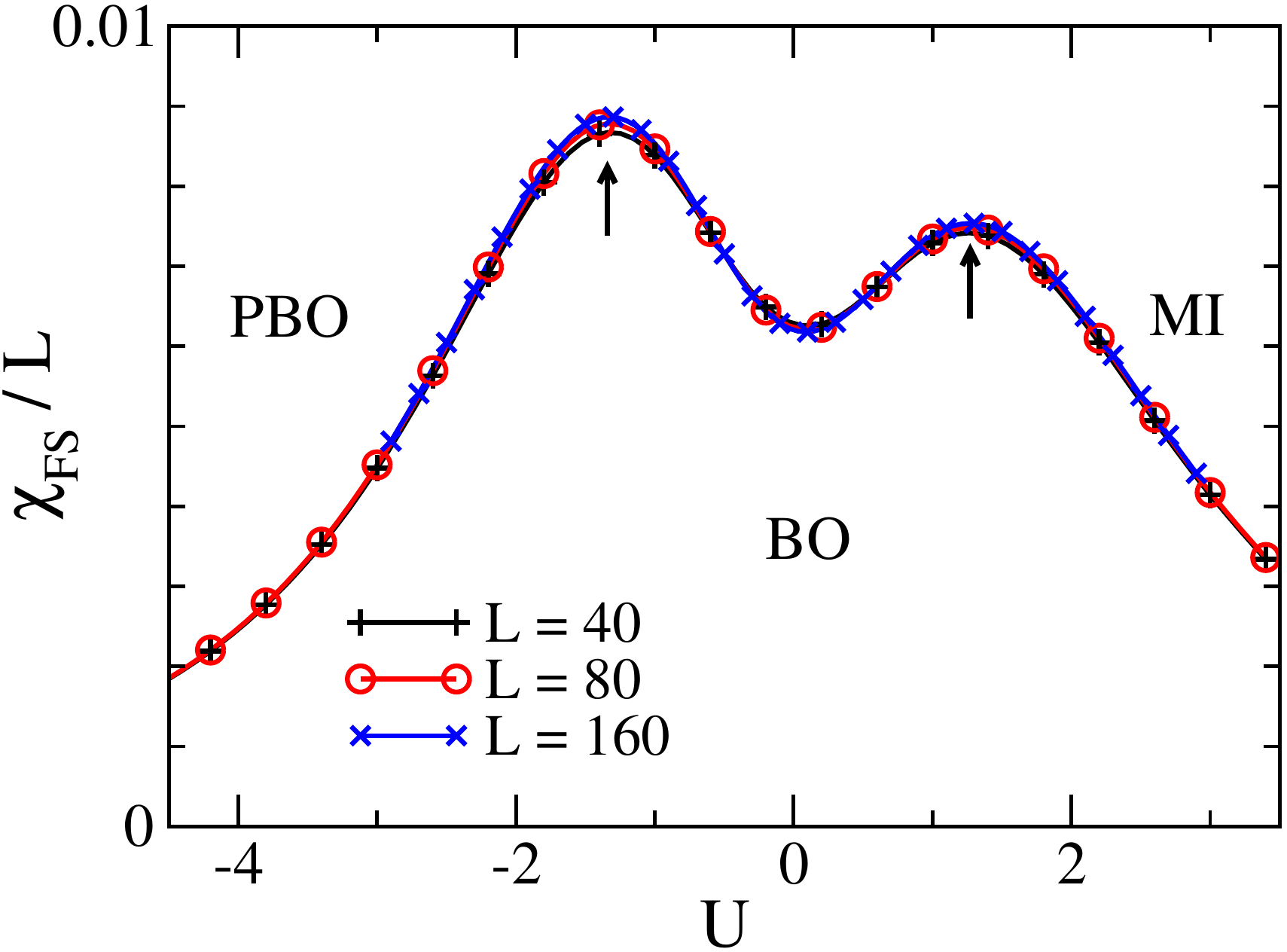}
  \caption{(Color online)Fidelity susceptibility $\chi_{FS}$ as function of the
  interaction strength $U$ ($\rho=1$, $t_2=0.2$) for $L=40$, $80$ and $160$.}
  \label{fig:fs}
\end{figure}
We may further characterize the position of the phase cross-over by
local maximum of the fidelity susceptibility ~\cite{gu2010} which is defined as

\begin{align}
\chi_{FS}(U) = \lim_{U-U' \to 0} \frac{-2 \ln |\langle \Psi_0(U) |
\Psi_0(U') \rangle| }{(U-U')^2} \,,
\end{align}
with the ground-state wave function $|\Psi_0\rangle$. While the phase
transitions are often characterized by a peak diverging with the system
size $L$, here we observe a stable maximum as function of several
system sizes as shown in Fig.~\ref{fig:fs}.
 
\begin{figure}[!b]
  \centering
  \includegraphics[width=\linewidth]{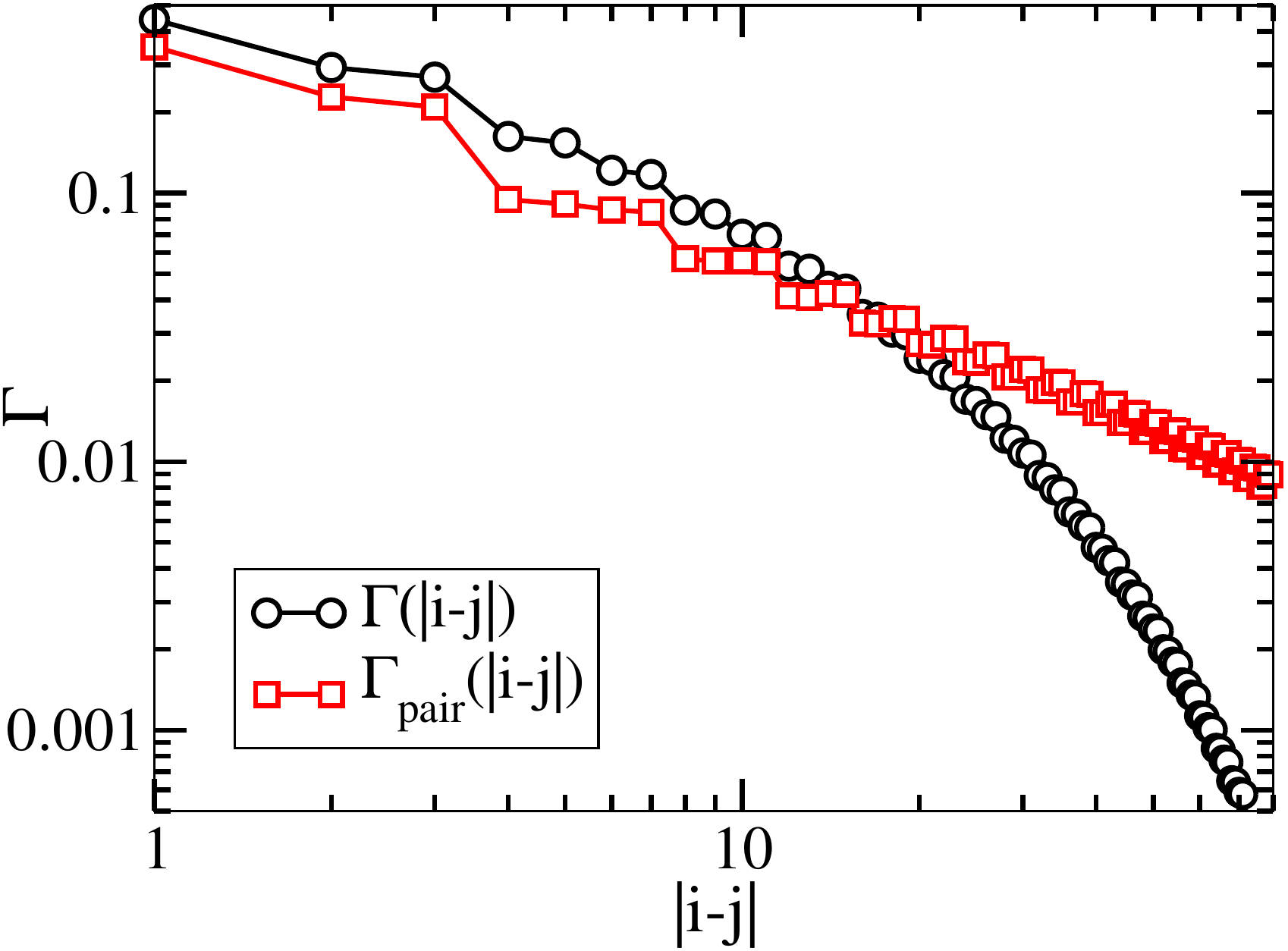}
    \caption{(Color online)Single particle correlation function $\Gamma(|i-j|)$ (black circles) and pair correlation functions 
    $\Gamma_{pair}(|i-j|)$ for $t_2=0.2$ are 
    plotted for $U=-4.0$ at $\rho=1.5$ 
    for a system of length $L=160$ (see text for details).}
    \label{fig:corr}
\end{figure}

The gapped BO phase which continues from the repulsive side for $\rho=1.5$ closes at a critical value {$U\sim-2.6$}. 
It is interesting to note that as the interaction becomes more and more attractive, the pair formation occurs and a PSF 
phase is stabilized for all densities around $\rho=1.0$~\cite{wessel,singh1}. 
The PSF and SF phases are separated by the red circles
as depicted in the phase diagram of Fig.~\ref{fig:fullpd1}. The signature of PSF phase can be obtained 
from the $\rho-\mu$ plot where the density jumps in steps of two particle at a time to 
minimize the energy. 
This is clearly visible in the insets of Fig.~\ref{fig:rhomub} 
where the densities for two different values of $U$ are plotted. For $U=-1.5$, the system is 
in a PSF phase in the region below the $\rho=1$ gapped phase and rest of the region is in the SF phase for 
incommensurate densities. 

\begin{figure}[!t]
  \centering
  \includegraphics[width=\linewidth]{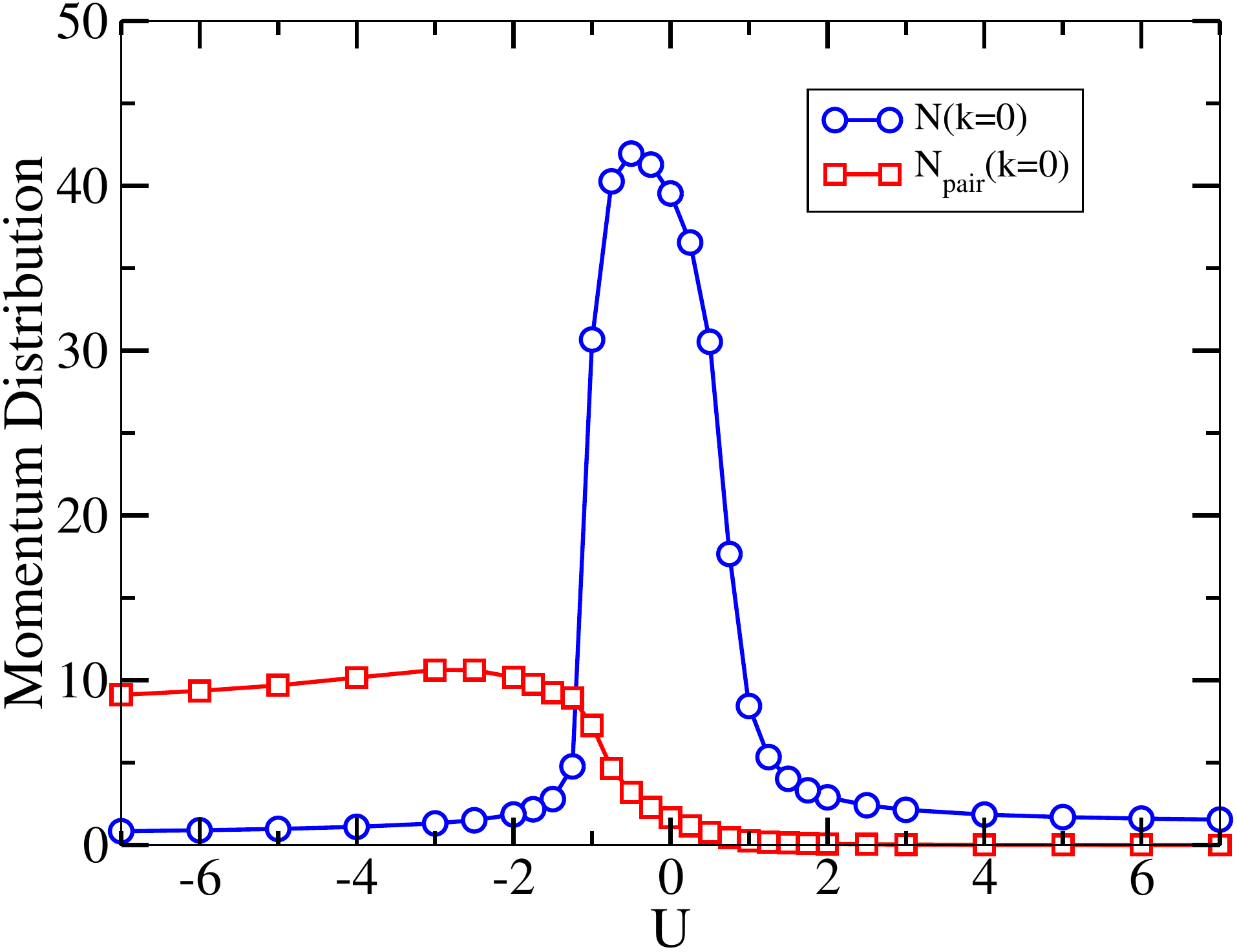}
    \caption{(Color online)The peak heights of $N(k=0)$ (blue circles) and $N_{pair}(k=0)$ (red squares) at $\rho=0.5$ using $L=160$ for $t_2=0.2$ show the transition from 
    PSF to SF and then to BO phases as discussed in the text.}
    \label{fig:mom}
\end{figure}
\begin{figure}[!b]
  \centering
  \includegraphics[width=\linewidth]{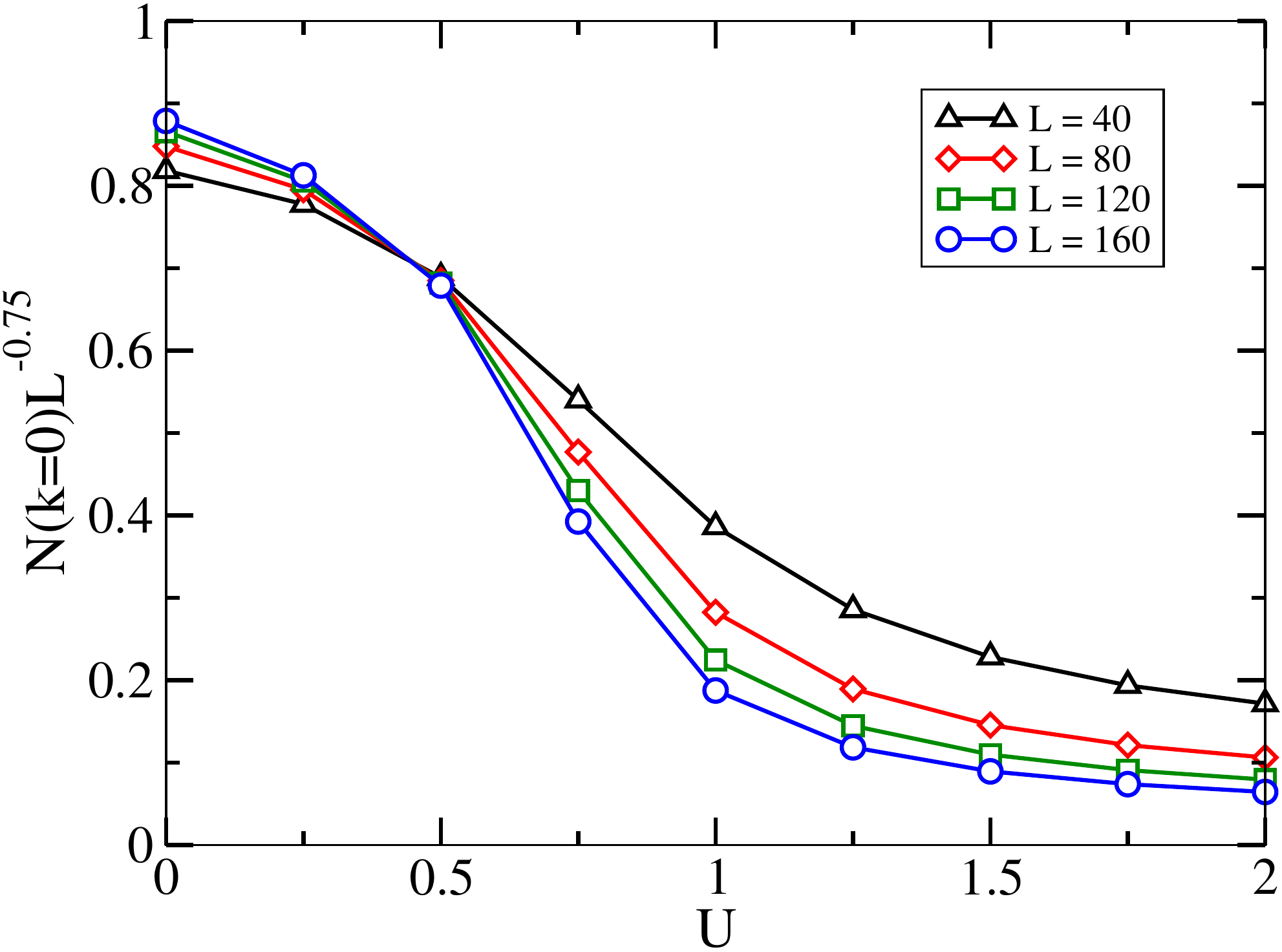}
    \caption{(Color online)Finite size scaling of $N(k=0)$ for $t_2=0.2$ shows that the curves for different lengths ($L=40,~80,~120,~160$) 
    intersect at the critical point $U\sim 0.44$ for the SF-MI transition at 
    $\rho=0.5$. }
    \label{fig:momscale}
\end{figure}
However, for $U=-4$ the system is in the PSF phase for all values of density except $\rho=1$. Apart from the $\rho-\mu$ curve, we compute the single particle and pair correlation functions and 
corresponding momentum distribution to 
confirm the existence of the PSF phase. Fig.~\ref{fig:corr} shows the behaviour of $\Gamma(i,j)=\langle a^\dagger_i a_j\rangle$(black circles) 
and $\Gamma_{pair}(i,j)=\langle {(a_i^\dagger})^2 (a_{j})^2 \rangle$(red squares) with respect to the 
distance $|i-j|$ for $U=-4$ and $\rho=1.5$ . It can be clearly seen that the single particle correlation function decays faster
where as the pair correlation function behaves like a power-law in the logarithmic 
scale which indicates the presence of the PSF phase. 

We also compute the momentum distribution function as
\begin{equation}
 N(k)=\frac{1}{L}\sum_{i,j}e^{ikr}\Gamma({i,j}),
 \label{eq:mom}
\end{equation}
to complement the SF phases. Where $\Gamma(i,j)=\langle a_i^\dagger a_j \rangle$($\langle {a_i^\dagger}^2 a_{j}^2 \rangle$) is the single particle(pair) correlation function. 
The peak heights of the momentum distribution function $N(k=0)$ for single particle and pairs are plotted against $U$ 
in Fig.~\ref{fig:mom} for a cut through the phase diagram of Fig.~\ref{fig:fullpd1} which corresponds to $\rho=0.5$. This shows that for large attractive interaction $N_{pair}(k=0)$ is dominant indicating the 
PSF phase. As the value 
of $U$ becomes less attractive then the value of $N_{pair}(k=0)$ decreases and $N(k=0)$ increases showing the signatures of the SF phase. The SF phase continues till the critical point for SF-BO 
transition on the repulsive side where both the momentum distribution functions are extremely small.
The SF-BO transitions are of Berezinskii-Kosterlitz-Thouless(BKT) type transition which can be seen from the smooth opening up of the gap in Fig.~\ref{fig:fullpd1}. 
The transition points can be accurately obtained by performing a finite size scaling of the single particle momentum distribution function 
which varies as $N(k=0) \propto ~ L^{1-\frac{1}{2K}}$~\cite{dhar}, where 
$K=2$ is the Luttinger parameter. In Fig.~\ref{fig:momscale} we plot $N(k=0)L^{-3/4}$ for different lengths($L=80,~120,~160$) and all the curves intersect at the critical point $U\sim0.44$.
\begin{figure}[!b]
  \centering
  \includegraphics[width=\linewidth]{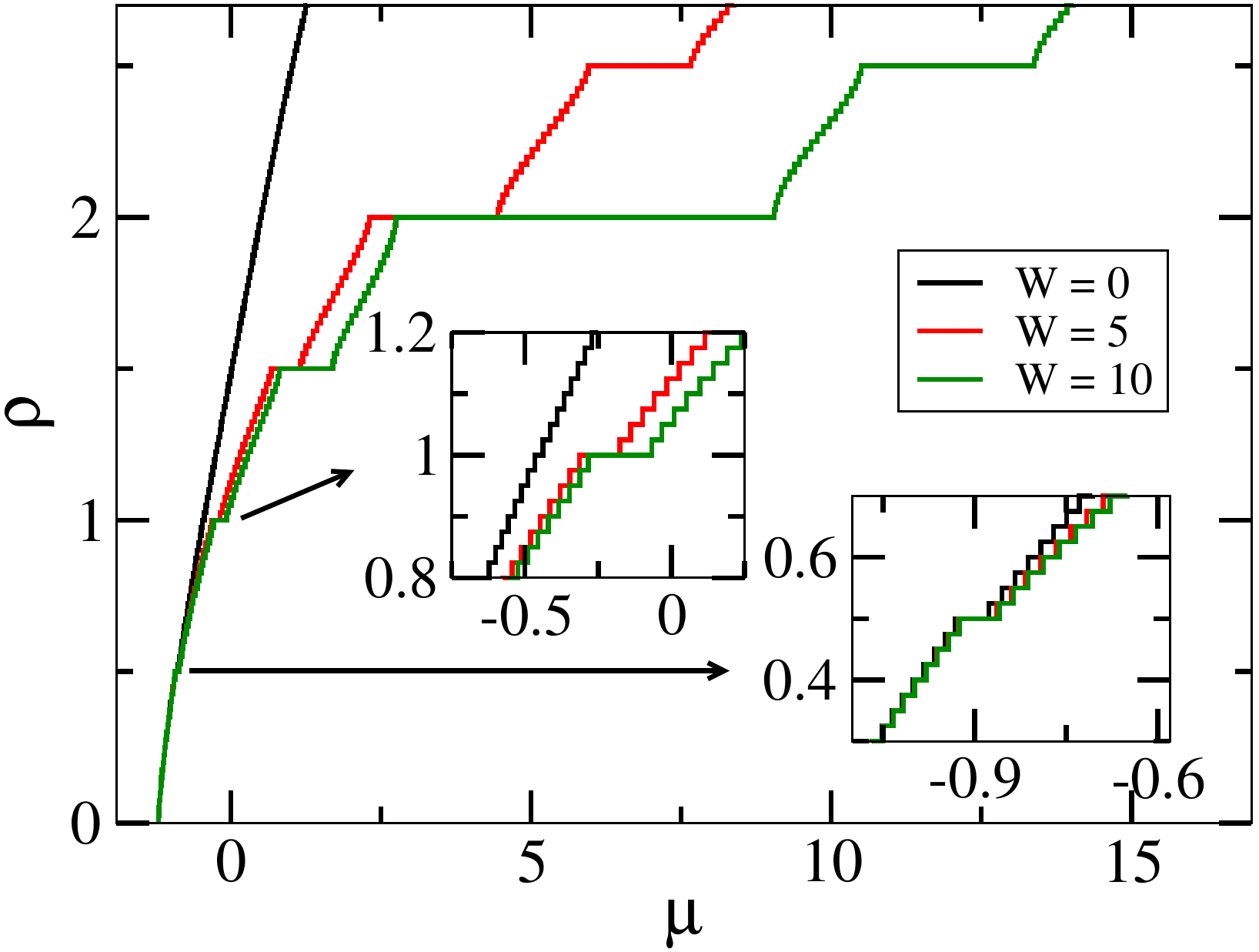}
    \caption{(Color online)Behaviour of $\rho$ with respect to $\mu$ for $U = 1$, $t_2=0.2$ and 
    $W=0$(black curve), $W=5$(red curve) and $W=10$(green curve) for the case of softcore bosons. Here maximum number of particle per site is taken to be 5. 
    Inset shows the zoomed in view of $\rho = 1$ and $0.5$ regions.}
    \label{fig:soft_rm}
\end{figure}

At this point we analyse the system without the three-body hardcore constraint to see the effect finite three-body interactions $(W/6) \sum_i n_i (n_i-1)(n_i-2) $ in the presence of the double-well potentials. 
As already mentioned before, for softcore bosons and vanishing two-body interaction i.e. $U=0$, the system is a gapless superfluid for any value of hopping dimerization which is 
in contrast to the case of three-body constrained bosons. To see the effect of finite $W$ we analyse the $\rho$ vs. $\mu$ plot for different values of $W=0,~5$ and $10$ considering $L=40$ 
with small onsite interaction $U=1$, which is shown in Fig~\ref{fig:soft_rm}. Here we can 
see that for $W=0$ the system is in the SF phase for integer and half integer densities except at half filling where there exists a small plateau. This is consistent with the result 
obtained in Ref~\cite{Fleischhauer2008}. However, as the value of $W$ increases the plateaus at other integer and half integer densities appear indicating various gapped phases. The plateau lenghts 
increase in size with increase in $W$. Therefore, we would like to highlight that although the three-body hardcore constraint is essential to stabilise the gapped phases on the attractive $U$ side, it is 
not that crucial for the repulsive $U$ case.

\subsection{Finite U and $t_2=0.6$ case}
\begin{figure}[!b]
  \centering
  \includegraphics[width=\linewidth]{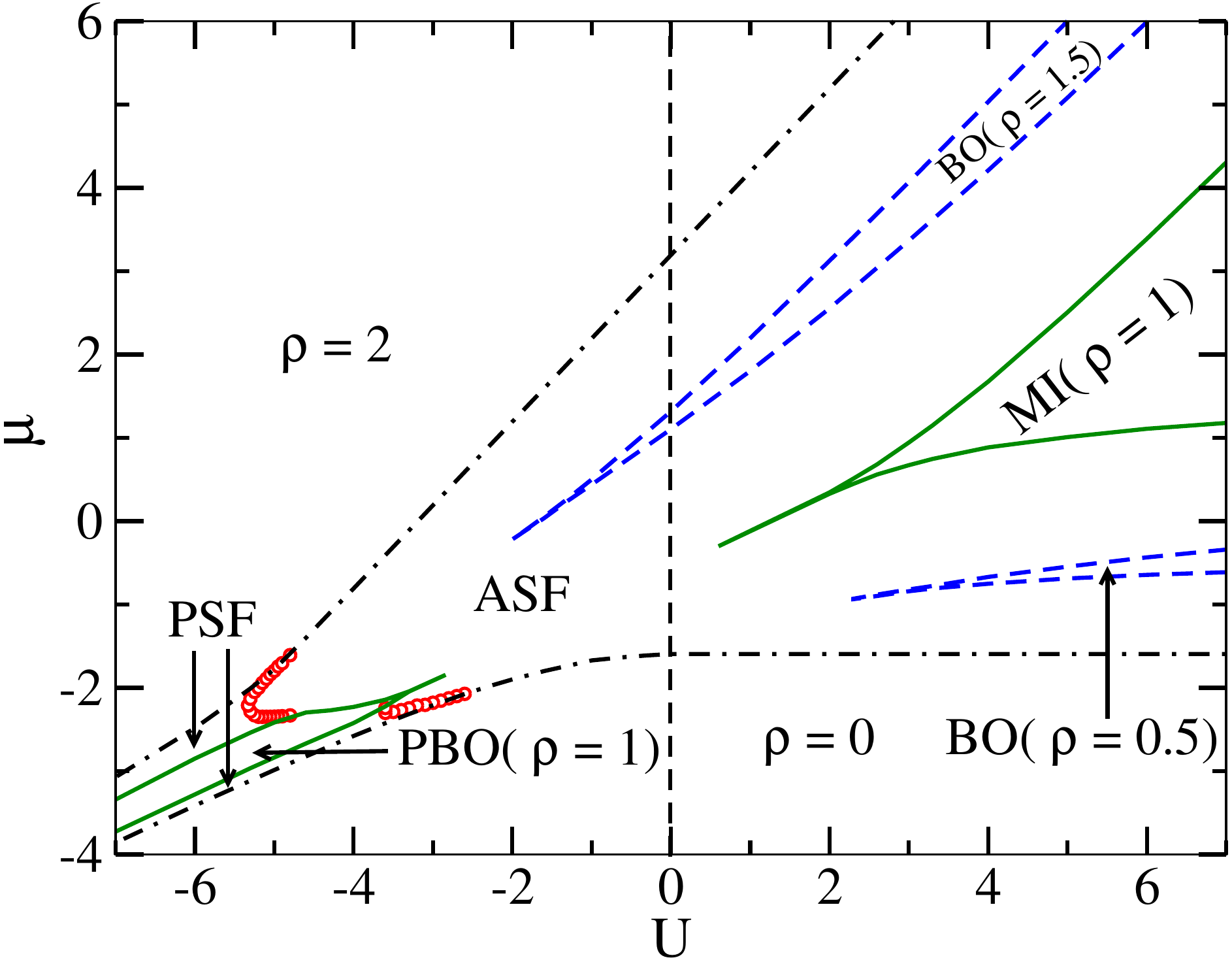}
    \caption{(Color online)Phase diagrams for $t_2=0.6$. The region bounded by the continuous green curves are the gapped phases at $\rho=1$ and 
    the regions bounded by the blue dashed curves are the gapped bond order phases at $\rho=0.5$ (upper) and $1.5$ (lower). On the negative $U$ side the PSF is separated from the SF phase by the red 
    circles. The black dot-dashed lines represent the empty and full states. }
    \label{fig:fullpd2}
\end{figure}
\begin{figure}[!t]
  \centering
  \includegraphics[width=\linewidth]{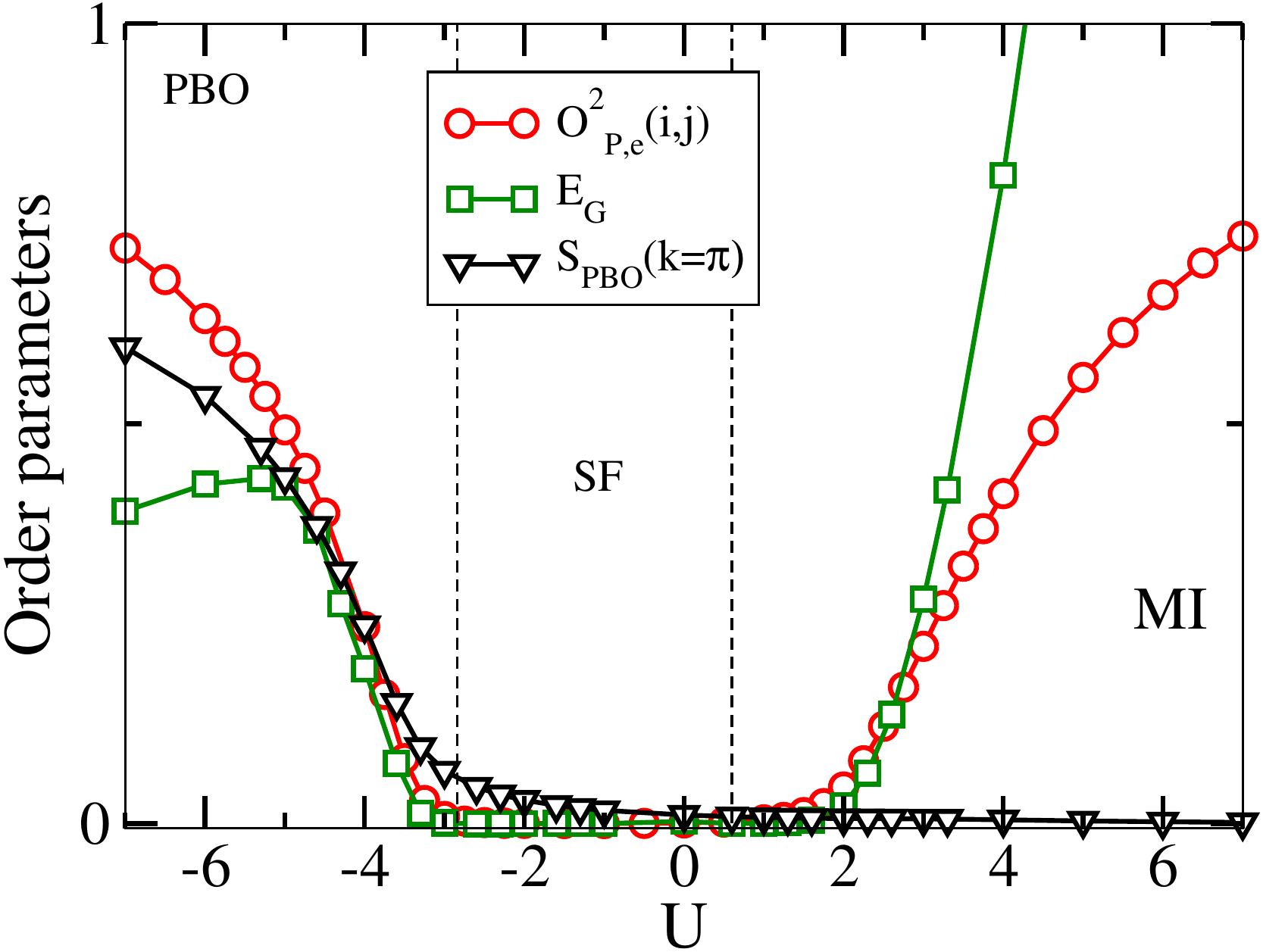}
    \caption{(Color online)Order parameters depicting different phases for $t_2=0.6$. The red circles, green squares and black down triangles show the 
    values of $O_{P,e}^2({i,j})$, $E_G(L)$ and $S_{PBO}(k=\pi)$ respectively for different values of $U$.  }
    \label{fig:order}
\end{figure}

After analyzing the phase diagram for the $t_2=0.2$ case where the effect of staggered hopping is large, we repeat the calculation for another cut through the phase diagram of Fig.~\ref{fig:nonint} at 
$t_2=0.6$. The motivation to consider $t_2=0.6$ lies in the fact that there is no gap at $\rho=1$ for $U=0$ as depicted in Fig.~\ref{fig:nonint} and it will be interesting to see how the system evolves 
by moving away from this limit.  A similar analysis along the line of $t_2=0.2$ case leads to the phase diagram as shown in Fig.~\ref{fig:fullpd2}. 
It can be seen that the overall picture of the phase diagram is similar to that of $t_2=0.2$ for $\rho=0.5$ and $\rho=1.5$. 
However, interesting thing to note that there are {clear phase transitions from 
the SF phase to the MI phase on the repulsive side of $U$ and to a PBO phase on the attractive side of $U$}. 
These signatures can be clearly seen from various order parameters plotted in Fig.~\ref{fig:order}. It can be seen that the PBO structure factor, 
gap and the parity order(see Sec.~IIID for detail) remain finite in the gapped MI and PBO phases where as they vanish in the SF phase.



\subsection{Parity order}

\begin{figure}[!b]
  \centering
  \includegraphics[width=\linewidth]{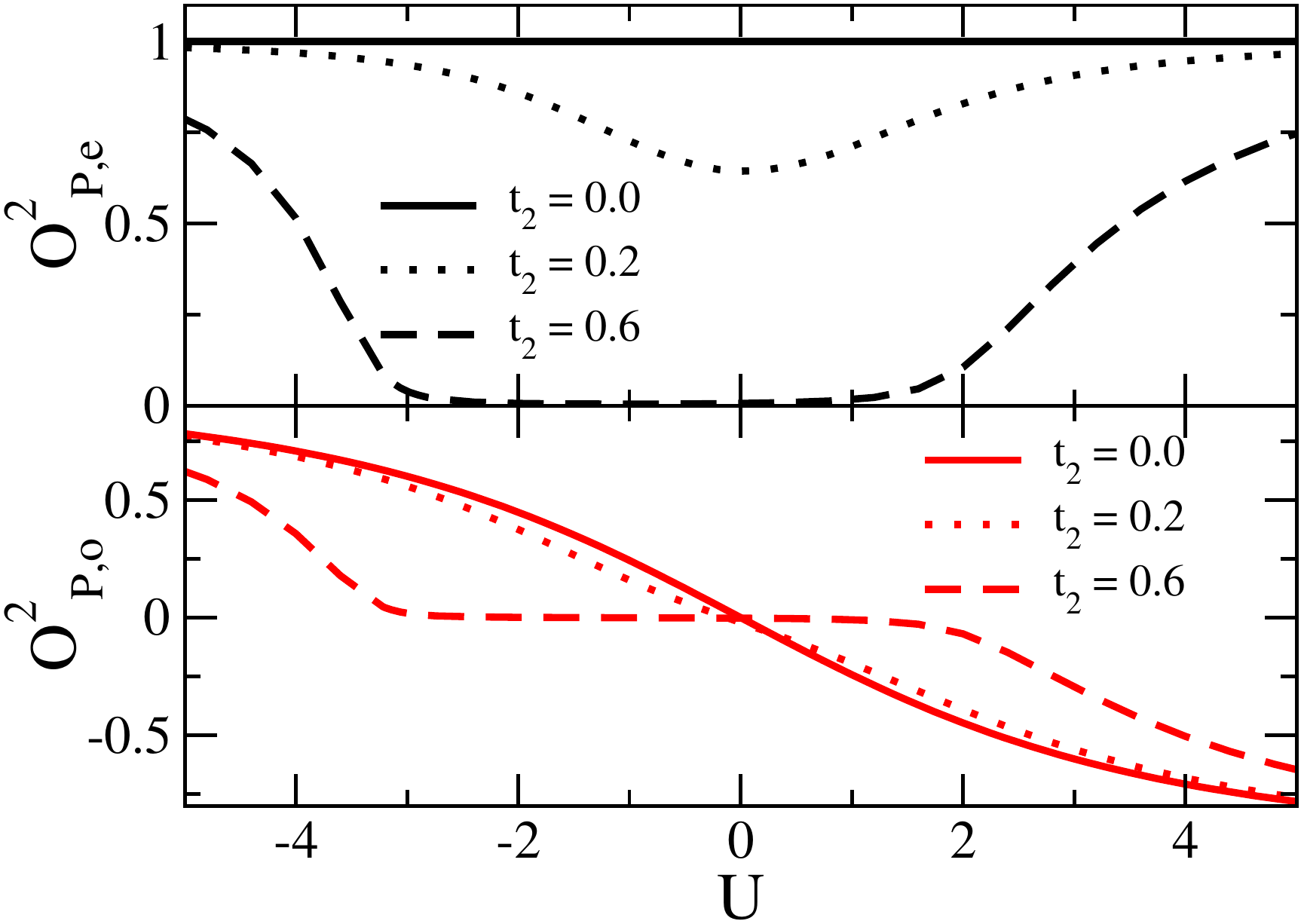}
    \caption{(Color online)Parity order $O_{P}^2(i,j)$ for odd and even distances $|i-j|\gg 1$ are plotted for different values of $t_2$ 
    with respect to $U$. $t_2=0$ lines 
    correspond to the analytical results as discussed in the main text, $t_2=0.2$ and $t_2=0.6$ curves depict 
    the data obtained for $L=160$ sites.  }
    \label{fig:n1_po}
\end{figure}
Another physical quantity of interest which can be directly accessed in the state-of-the art experiment~\cite{Bakr,Kuhr} is the parity order parameter which is defined as 
 \begin{equation}
 O_{P}^2({i,j})= \langle  e^{i\sum_{i<k<j}\pi n_k }\rangle
 \label{eq:parity}
\end{equation}

\begin{figure}[!t]
  \centering
  \includegraphics[width=\linewidth]{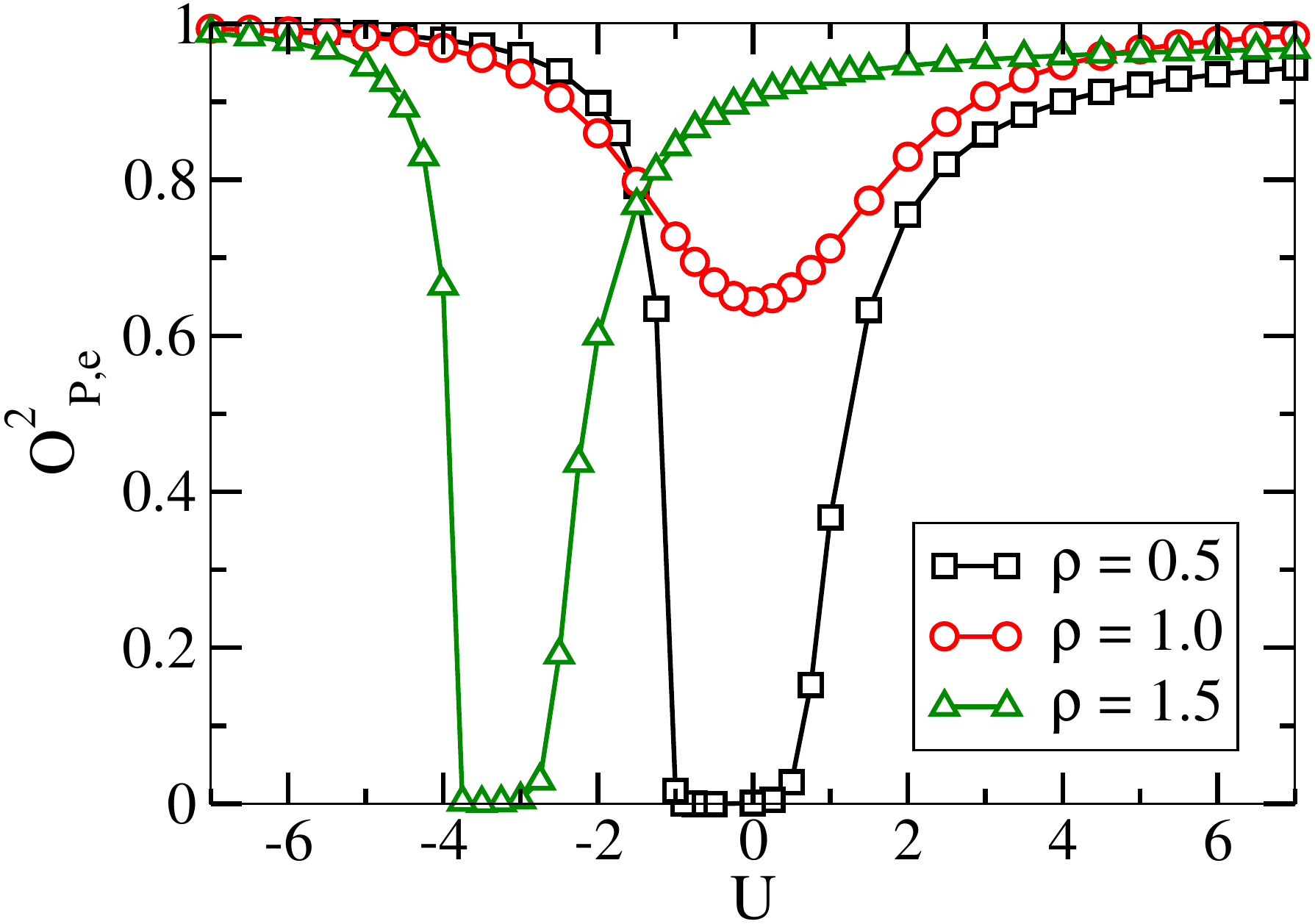}
    \caption{(Color online)$O_{P,e}^2$ at $\rho=0.5,~1.0$ and $1.5$ for $t_2=0.2$ is plotted for different values of $U$.  
     }
    \label{fig:parity}
\end{figure}

To complement our findings we compute $O_{P}^2({i,j})$  
which is finite in the MI phase due to particle-hole excitations. For small $t_2\to 0$ 
we may understand the emergence of parity order from the properties of isolated double wells as discussed in Sec.~\ref{sec:doublewell}. 
For a ground-state at $\rho=1$ given by a product of
$|\psi_1\rangle$, one easily estimates the parity order to be exactly
$O_{P,e} = 1$ on even distances $|i-j|$. For odd distances, however, one
observes an interesting dependence of the parity order on the
interaction strength, $O_{P,o} = -\frac{U}{\sqrt{16 t_1^2 + U^2}}$. 
We plot the odd and even distance parity orders as $O_{P,e}$ and $O_{P,o}$ respectively with respect to $U$ for different values of $t_2$ in Fig.~\ref{fig:n1_po}. 
The black and red curves correspond to $O_{P,e}$ and $O_{P,o}$ respectively.
The solid lines correspond to the 
limit of isolated double wells i.e for $t_2=0$. The dotted lines are the $t_2=0.2$ and the dashed curves are for $t_2=0.6$. 
It is very clear from this figure that the odd and even distance 
parities show two different behaviour and the parity order is 
finite in the MI, BO and PBO phases where as it is zero in the SF phase. 
Analogously one also finds for the gapped phases at half filling a finite oscillating parity order which is $0$ for even and $\pm 1$ for odd distances.
The parity order parameter also vanishes in the SF phase. This can be seen from the Figure.~\ref{fig:parity} 
where $O_{P,e}^2$ is plotted as a function of $U$ for different fillings.
\section{Conclusions}
We investigated the ground state phase diagram of a system of three-body constrained bosons in a double well optical lattice. By analysing the competition between the 
dimerized hopping and onsite interactions we obtained the complete phase diagram both in the regime of attractive as well as repulsive interactions. 
The phase diagram exhibits various gapped phases such as the BO and MI phases at commensurate densities. At unit filling and small hopping ratios 
the system exhibits a BO phase of pairs in the attractive interaction regime which we call the PBO phase and 
there exists smooth crossover from MI-BO~(BO-PBO) on the repulsive~(attractive) 
side of the phase diagram. For large values of the hopping ratios, the gapped phases melt for small values of interactions and there exist an 
intermediate SF phase. The appropriate finite size scaling 
shows that the superfluid to gapped phase transitions are of BKT type. 
The findings presented in this work addresses an interesting problem which involves the physics of strongly correlated bosons 
in a double well optical lattice both in attractive and repulsive regime. 
As the double well optical lattices have already been created and manipulated using cold atoms, it will be possible to observe these 
phases in the experiments.

As mentioned before, this kind of double well optical lattice resembles the topological SSH model discussed in the context of 
solitons in polyacetylene, which possess two types of dimerizations depending on the hopping ratios. The SSH model exhibits topological 
phase transition from a trivial to non trivial phase through a gapless point. The non trivial phase is characterized by zero 
energy edge modes. The topological aspects of this model have been analysed recently in various contexts~\cite{Grusdt2013,Gonzalez2018,Gonzalez2018b,Yoshida2014,Yoshida2018}. 
One of the interesting phenomenon which signals these topological
phase transition is the Thouless charge pumping mechanism~\cite{Schweizer2016,Takahashi2016pumping,Kraus2012pumping,Nakagawa2018}. In the present scenario the PBO phase consists of
hardcore boson pairs and it is in principle 
possible to map the system to an effective SSH model for the spinless fermions and study the topological features in the context of Rice-Mele model~\cite{Rice1982,Hayward2018,ssh2}.

\begin{acknowledgments}
The  computational simulations were carried out using the Param-Ishan HPC facility at Indian Institute of Technology - Guwahati, India. 
S.G. acknowledges support by the Swiss National Science Foundation under Division II. 
T.M. acknowledges DST-SERB for the early career grant through Project No.  ECR/2017/001069.
\end{acknowledgments}



\bibliography{references}

\end{document}